\documentclass[preprint,12pt,3p]{elsarticle}

\usepackage[utf8]{inputenc}
\usepackage{CJKutf8}

\usepackage{graphicx}

\usepackage{amssymb}
\usepackage{eso-pic}

\journal{Icarus}

\begin{document}
\newcommand{\NTnamea}{2013~VX$_{30}$} 
\newcommand{\NTnameb}{2014~UU$_{240}$} 
\newcommand\aj{{Astronomical Journal}}
\newcommand\apj{{Astrophysical Journal}}
\newcommand\apjl{{Astrophysical Journal Letters}}
\newcommand\apjs{{Astrophysical Journal Supplement Series}}
\newcommand\aap{{Astronomy and Astrophysics}}
\newcommand\icarus{{Icarus}}
\newcommand\mnras{{Monthly Notices of the Royal Astronomical Society }}
\newcommand\pasj{{Publications of the Astronomical Society of Japan }}
\newcommand\nat{{Nature}}

\begin{frontmatter}

\title{Evidence for Color Dichotomy in the Primordial Neptunian Trojan Population} 

\author[affil1]{Hsing~Wen~Lin (\begin{CJK*}{UTF8}{bkai}
林省文\end{CJK*})}
\address[affil1]{Department of Physics, University of Michigan, Ann Arbor, Michigan 48109 USA}

\ead{hsingwel@umich.edu}

\author[affil1,affil2]{David W. Gerdes}
\address[affil2]{Department of Astronomy, University of Michigan, Ann Arbor, Michigan 48109 USA}
\ead{gerdes@umich.edu}

\author[affil1,note1]{Stephanie J. Hamilton}
\ead{sjhamil@umich.edu}
\fntext[note1]{NSF Graduate Fellow}
\author[affil1,affil2]{Fred C. Adams}
\ead{fca@umich.edu}

\author[affil3]{Gary M. Bernstein}

\author[affil3]{Masao Sako}

\author[affil3]{Pedro Bernadinelli}

\author[affil4]{Douglas Tucker}
\author[affil4]{Sahar Allam}

\author[affil2,note1]{Juliette C. Becker}

\author[affil1]{Tali Khain} 

\author[affil2, note1]{Larissa Markwardt}

\author[affil1]{Kyle Franson}

\address[affil3]{Department of Physics and Astronomy, University of Pennsylvania, Philadelphia, PA 19104, USA}
\address[affil4]{Fermi National Accelerator Laboratory, P. O. Box 500, Batavia, IL 60510, USA}


\author[5]{T.~M.~C.~Abbott}
\author[affil4]{J.~Annis}
\author[6]{S.~Avila}
\author[7]{D.~Brooks}
\author[8,9]{A.~Carnero~Rosell}
\author[10,11]{M.~Carrasco~Kind}
\author[12]{C.~E.~Cunha}
\author[4]{C.~B.~D'Andrea}
\author[8,9]{L.~N.~da Costa}
\author[13]{J.~De~Vicente}
\author[7]{P.~Doel}
\author[14,15]{T.~F.~Eifler}
\author[affil4]{B.~Flaugher}
\author[16]{J.~Garc\'ia-Bellido}
\author[17]{D.~L.~Hollowood}
\author[18,19]{Klaus Honscheid}
\author[20]{D.~J.~James}
\author[21]{K.~Kuehn}
\author[affil3]{N.~Kuropatkin}
\author[8,9]{M.~A.~G.~Maia}
\author[22]{J.~L.~Marshall}
\author[23,24]{R.~Miquel}
\author[15]{A.~A.~Plazas}
\author[25]{A.~K.~Romer}
\author[13]{E.~Sanchez}
\author[affil4]{V.~Scarpine}
\author[13]{I.~Sevilla-Noarbe}
\author[26]{M.~Smith}
\author[5]{R.~C.~Smith}
\author[27]{M.~Soares-Santos}
\author[28,8]{F.~Sobreira}
\author[29]{E.~Suchyta}
\author[affil1]{G.~Tarle}
\author[5]{A.~R.~Walker}
\author[affil4]{W.~Wester}

\address[5]{Cerro Tololo Inter-American Observatory, National Optical Astronomy Observatory, Casilla 603, La Serena, Chile}
\address[6]{Institute of Cosmology \& Gravitation, University of Portsmouth, Portsmouth, PO1 3FX, UK}
\address[7]{Department of Physics \& Astronomy, University College London, Gower Street, London, WC1E 6BT, UK}
\address[8]{Laborat\'orio Interinstitucional de e-Astronomia - LIneA, Rua Gal. Jos\'e Cristino 77, Rio de Janeiro, RJ - 20921-400, Brazil}
\address[9]{Observat\'orio Nacional, Rua Gal. Jos\'e Cristino 77, Rio de Janeiro, RJ - 20921-400, Brazil}
\address[10]{Department of Astronomy, University of Illinois at Urbana-Champaign, 1002 W. Green Street, Urbana, IL 61801, USA}
\address[11]{National Center for Supercomputing Applications, 1205 West Clark St., Urbana, IL 61801, USA}
\address[12]{Kavli Institute for Particle Astrophysics \& Cosmology, P. O. Box 2450, Stanford University, Stanford, CA 94305, USA}
\address[13]{Centro de Investigaciones Energ\'eticas, Medioambientales y Tecnol\'ogicas (CIEMAT), Madrid, Spain}
\address[14]{Department of Astronomy/Steward Observatory, 933 North Cherry Avenue, Tucson, AZ 85721-0065, USA}
\address[15]{Jet Propulsion Laboratory, California Institute of Technology, 4800 Oak Grove Dr., Pasadena, CA 91109, USA}
\address[16]{Instituto de Fisica Teorica UAM/CSIC, Universidad Autonoma de Madrid, 28049 Madrid, Spain}
\address[17]{Santa Cruz Institute for Particle Physics, Santa Cruz, CA 95064, USA}
\address[18]{Center for Cosmology and Astro-Particle Physics, The Ohio State University, Columbus, OH 43210, USA}
\address[19]{Department of Physics, The Ohio State University, Columbus, OH 43210, USA}
\address[20]{Harvard-Smithsonian Center for Astrophysics, Cambridge, MA 02138, USA}
\address[21]{Australian Astronomical Observatory, North Ryde, NSW 2113, Australia}
\address[22]{George P. and Cynthia Woods Mitchell Institute for Fundamental Physics and Astronomy, and Department of Physics and Astronomy, Texas A\&M University, College Station, TX 77843,  USA}
\address[23]{Instituci\'o Catalana de Recerca i Estudis Avan\c{c}ats, E-08010 Barcelona, Spain}
\address[24]{Institut de F\'{\i}sica d'Altes Energies (IFAE), The Barcelona Institute of Science and Technology, Campus UAB, 08193 Bellaterra (Barcelona) Spain}
\address[25]{Department of Physics and Astronomy, Pevensey Building, University of Sussex, Brighton, BN1 9QH, UK}
\address[26]{School of Physics and Astronomy, University of Southampton,  Southampton, SO17 1BJ, UK}
\address[27]{Brandeis University, Physics Department, 415 South Street, Waltham MA 02453}
\address[28]{Instituto de F\'isica Gleb Wataghin, Universidade Estadual de Campinas, 13083-859, Campinas, SP, Brazil}
\address[29]{Computer Science and Mathematics Division, Oak Ridge National Laboratory, Oak Ridge, TN 37831}


\AddToShipoutPictureBG*{%
  \AtPageUpperLeft{%
    \hspace{0.7\paperwidth}%
    \raisebox{-4\baselineskip}{%
    
      \makebox[0pt][l]{\textnormal{DES 2018-0359}}   
}}}%

\AddToShipoutPictureBG*{%
 \AtPageUpperLeft{%
    \hspace{0.7\paperwidth}%
    \raisebox{-5\baselineskip}{%
      \makebox[0pt][l]{\textnormal{FERMILAB-PUB-18-294-AE}}   
}}}%

\AddToShipoutPictureBG*{%
 \AtPageUpperLeft{%
    \hspace{0.7\paperwidth}%
    \raisebox{-6\baselineskip}{%
      \makebox[0pt][l]{\textnormal{Draft version 2.1, \today}}
}}}%

\begin{abstract}
In the current model of early Solar System evolution, \textcolor{black}{the stable members of the Jovian and Neptunian Trojan populations were captured into resonance from the leftover reservoir of planetesimals during the outward migration of the giant planets.} As a result, both Jovian and Neptunian Trojans share a common origin with the primordial disk population, whose other surviving members constitute today's trans-Neptunian object (TNO) populations. The cold (low inclination and small eccentricity) classical TNOs are ultra-red, while the dynamically excited ``hot" (high inclination and larger eccentricity) population of TNOs contains a mixture of ultra-red and blue objects. In contrast, Jovian and Neptunian Trojans are observed to be blue. While the absence of  ultra-red Jovian Trojans can be readily explained by the sublimation of volatile material from their surfaces due to the high flux of solar radiation at 5~AU, the lack of ultra-red Neptunian Trojans presents both a puzzle and a challenge to formation models. In this work we report the discovery by the Dark Energy Survey (DES) of two new dynamically stable L4 Neptunian Trojans, \NTnamea\ and \NTnameb, both with inclinations $i>30^{\circ}$, making them the highest-inclination known stable Neptunian Trojans. We have measured the colors of these and three other dynamically stable Neptunian Trojans previously observed by DES, and find that \NTnamea\ is ultra-red, the first such Neptunian Trojan in its class. As such, \NTnamea\ may be a ``missing link" between the Trojan and TNO populations. Using a simulation of the DES TNO detection efficiency, we find that there are 162 $\pm$ 73 Trojans with $H_r < 10$ at the L4 Lagrange point of Neptune. Moreover, the blue-to-red Neptunian Trojan population ratio should be higher than \textcolor{black}{17:1}. Based on this result, we discuss the possible origin of the ultra-red Neptunian Trojan population and its implications for the formation history of Neptunian Trojans.

\end{abstract}


\end{frontmatter}


\section{Introduction}
\label{sec:intro}

The Trojan asteroids reside in asymmetric 1:1 mean-motion resonances with planets, librating around the planet's L4 (leading) and L5 (trailing) Lagrange points. Unlike the symmetric librators, which have horseshoe co-orbital motion with the planet and are usually not dynamically  stable in the long-term \citep{Brasser2004:Transient, Marcos2012}, the Trojans can be stable for the entire age of the Solar System \citep{Nesvorny2002, Marzari2003}. Jupiter holds by far the largest number of known Trojan asteroids, with Neptune having the second-most. Neptune has roughly two dozen known Trojans, but many previous studies suggest that the Neptunian Trojan population should be even larger than that of Jupiter \citep{Alexandersen2016, Chiang2003, Lin2018, Sheppard2010:size_NT, Sheppard2010:L5NT}.

\textcolor{black}{Both the Jovian and Neptunian Trojan populations are thought to have arisen from capture during the planetary migration or orbital damping stage after the dispersal of the protoplanetary disk rather than via {\em in-situ} formation.} Several arguments suggest this scenario. First, both the Jovian and Neptunian Trojan populations have wide inclination distributions rather than the nearly planar distribution of the original disk of planetesimals \citep{Sheppard2006}. Additionally, numerical simulations show that the very first Trojans (those that formed {\em in-situ} prior to planetary migration) were destabilized and eventually lost during planetary migration \citep{Chen2016, Chiang2005, Kortenkamp2004}. In the context of the Nice model \citep{Gomes2005, Tsiganis2005}, the Jovian Trojans could have formed in more distant regions (such as the trans-Neptunian region) and been captured into co-orbital motion with Jupiter during planetary migration when the dynamics of the Trojan region were completely chaotic \citep{Morbidelli2005}. This process, generally known as chaotic capture, can operate similarly to capture the Neptunian Trojans \citep{Nesvorny2009}. However, \citet{Lykawka2010:capture_NT} have argued that high inclination ($i>25^\circ$) orbits of Neptunian Trojans are unlikely to be produced by fast migration. As a result, scenarios with slower migration rates should be considered. 

In any case, all of the current mechanisms for planetary migration and capture indicate that the source of primordial Jovian and Neptunian Trojans --- that is, the Trojans that have librated in stable orbits since the end of planetary migration --- should be the trans-Neptunian region. As a result, the trans-Neptunian objects (TNOs) should share some common properties with Trojans. As one piece of evidence in support of this scenario, many studies have demonstrated the similar size distributions between Trojans and the ``hot'' population of TNOs \citep{Morbidelli2009, Fraser2014, Yoshida2017}. 

This paper addresses another key physical property of Neptunian Trojans: their colors. The observed TNOs display a well-known and wide range of colors, from blue to ultra-red \citep{Hainaut2012, Lacerda2014, Peixinho2015, Peixinho2012, Pike2017, Schwamb2018, Sheppard2012, Sheppard2010:color_ETNOs, Terai2018, Wong2017}. If the Jovian and Neptunian Trojans share a common source with the ``hot'' population of the TNOs, we would expect to observe the same wide color distribution. \textcolor{black}{However, Jovian Trojans, even thought have been found bi-modal in color \citep{Wong2016, Wong2015}, are mostly have spectra resembling either D- or P-type asteroids \citep{DeMeo2014, DeMeo2013}, which display relatively narrow region in color space \citep{Grav2012, Szabo2007}}. Moreover, recent observations of the Neptunian Trojans also indicate that the colors of Neptunian Trojans are quite similar to those of the Jovian Trojans \citep{Gerdes2016, Jewitt2018, Parker2013, Sheppard2012, Sheppard2006}. Neither of these populations contains any members with ultra-red colors. 

Color differences could arise due to the different distances of minor bodies from the Sun. Because Jupiter's orbit has a semi-major axis of only $a\approx5.12$ AU, the Jovian Trojans have blackbody surface temperatures $T\sim$ 125~K, which is much warmer than the TNOs ($T\sim$ 40 K). As a result, the observed color difference between Jovian Trojans and TNOs could be explained by the possible resurfacing of the inner population, i.e. the sublimation of volatile materials due to their elevated temperature \citep{Grundy2009, Wong2016}. However, the Neptunian Trojans have heliocentric distances ($d\sim$ 30 AU) and blackbody temperatures ($T\sim$ 50 K) similar to those of TNOs, and thermal-resurfacing processes are not expected to operate at the distance of Neptune's orbit. Resurfacing through collisions could be an alternative process to remove the ultra-red surfaces. However, the collision rate in the Neptunian Trojan swarms is dominated by very small bodies and such collisions remain essentially unobserved. \textcolor{black}{Moreover, although there is an collisional asteroid family in Jovian Trojans -- the Eurybates family \citep{Bro2011}, no such system have been found in Neptunian Trojan population yet.} In summary, thermal resurfacing processes cannot explain the lack of ultra-red Neptunian Trojans. Collisional resurfacing processes may contribute, but current observational evidence is lacking \citep{Jewitt2018}.

As outlined above, the observed similarity in colors between Jovian and Neptunian Trojans poses an interesting puzzle. It suggests that Jovian and Neptunian Trojans could have common sources or that they experience similar resurfacing processes. However, the color distributions of the Trojans differ from their plausible common-source population, namely the TNOs, and there is no known resurfacing process that performs equally well at the (widely different) heliocentric distances of Jupiter and Neptune. An obvious possible solution that sidesteps this problem would be to establish color differences between the populations of Jovian and Neptunian Trojans. More specifically, it would be significant to find ultra-red members of the Neptunian Trojan population. In this event, the color distribution of Neptunian Trojans would more closely resemble their putative common source, the TNOs, and the observed blue colors of Jovian Trojans would be explained by thermal resurfacing. Toward this end, this work reports colors for five dynamically stable Neptunian Trojans observed by the Dark Energy Survey (DES), including the first-ever dynamically stable ultra-red Neptunian Trojan. 

This paper is organized as follows: In  Section~\ref{sec:obs}, we introduce the discovery of two new Neptunian Trojans, \NTnamea\ and \NTnameb, by the Dark Energy Survey. In Section~\ref{sec:dyna}, we numerically integrate 1000 clones of these bodies to demonstrate their long-term stability. These results suggest that they belong to a long-term stable population, rather than a transient population temporarily captured into resonance. In Section~\ref{sec:phot}, we provide the multi-color photometry for the five Neptunian Trojans observed by DES. In Section~\ref{sec:pop}, we estimate the ratio of the blue and ultra-red Trojan populations using our five DES Trojans. In Section~\ref{sec:dis}, we discuss possible origin scenarios for ultra-red Neptunian Trojans. Finally, Section~\ref{sec:sum} provides a summary of our results and a discussion of their implications. 

\section{Discovery}
\label{sec:obs}

\begin{figure}
\centering
\includegraphics[width = 0.9\textwidth]{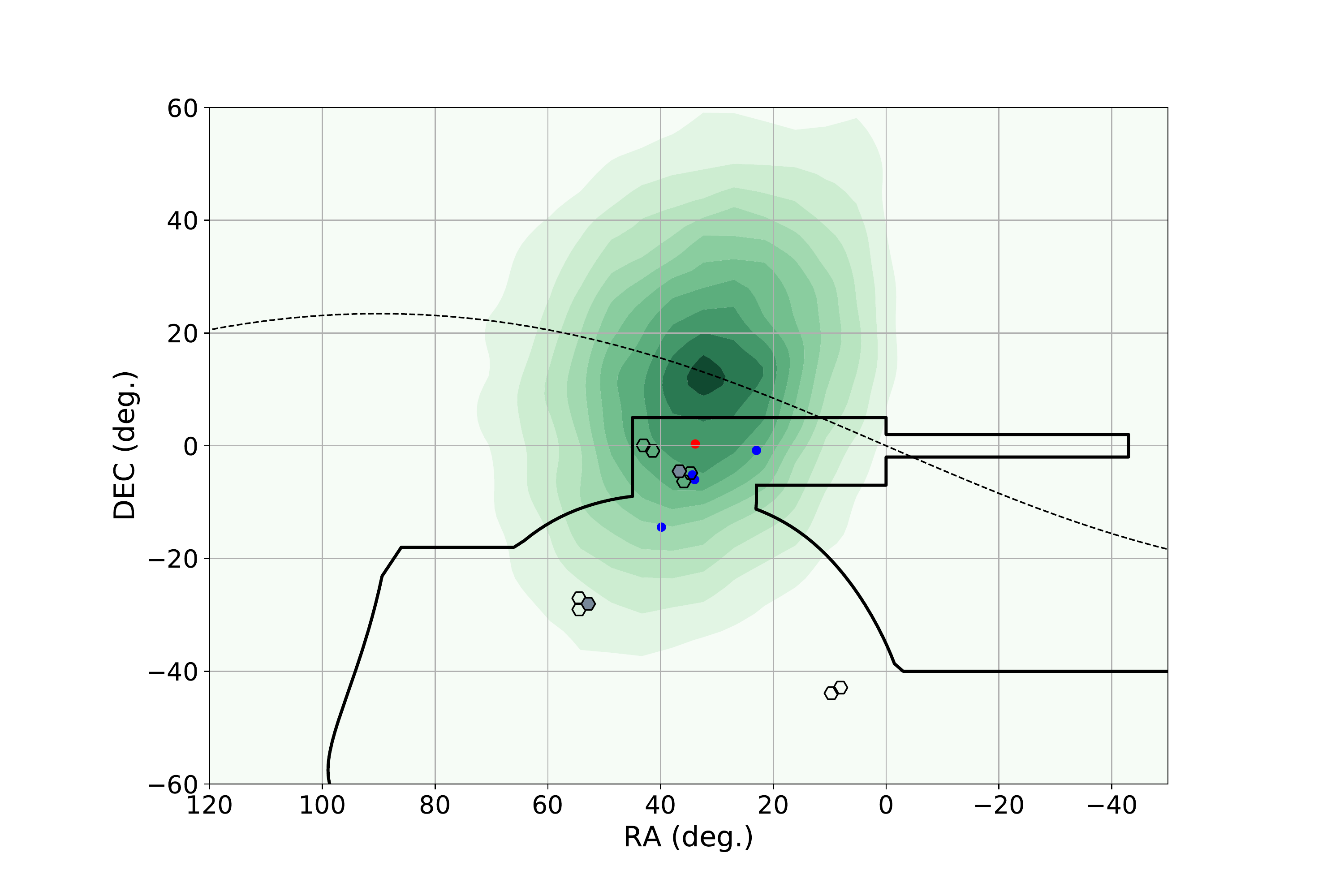}
\caption{The northern portion of the Dark Energy Survey footprint, superimposed on a density 
plot of the L4 Neptunian Trojan cloud modeled as described in Sec.~\ref{sec:pop}. The ecliptic is shown as a dashed line. The DES wide survey area is outlined in black. The ten supernova survey fields are shown
as small hexes, with the two deep fields shaded gray. The survey region encompasses
roughly 25\% of the L4 Trojans. The locations (on 2015 Jan.\ 1) of the five detected Trojans discussed in this work are shown as dots, with \NTnamea\ in ultra-red and the other four objects in blue. The two objects discovered in the deep supernova field \citep{Gerdes2016} are nearly superimposed at this scale.
\label{fig:NTrojans_DES}}
\end{figure}

The Dark Energy Survey (DES) is an optical survey being carried out with the Dark Energy Camera (DECam) \citep{DECam2015}, a 3 sq.\ deg.\ prime focus imager on the 4-meter Blanco telescope at the Cerro Tololo Inter-American Observatory in Chile. The survey footprint
is shown as the black outline in Figure~\ref{fig:NTrojans_DES}. It is evident from this figure that the DES is well-suited to 
the study of the L4 Neptunian Trojan cloud.

The DES was awarded 525 nights over 5 years, beginning in 2013. Each season's observing campaign spans 105 nights from roughly mid-August to mid-February. The scientific program consists of two simultaneous surveys. 
The \textit{supernova survey} \citep{Bernstein2012} (small hexes in Figure~\ref{fig:NTrojans_DES}) targets 10 individual DECam fields ($\approx 30$ sq.\ deg.\ altogether)  at approximately 6-day intervals throughout each DES observing campaign. Eight of these supernova fields are ``shallow", with successive images taken in the $griz$ bands over a roughly 15-minute interval. The remaining two supernova fields are ``deep", with stacked exposure sequences ranging from 10 minutes in the $g$-band to 60 minutes in the $z$-band. The \textit{wide survey} covers 5000 sq.\ deg.\ of the south galactic cap, but at a much sparser and less regular cadence. When the survey is 
completed, each wide survey tiling will have been observed approximately 10 times in each of the $grizY$ bands. 
The limiting  single-exposure depth in $r$-band wide-survey exposures is approximately 23.5; supernova survey exposures reach approximately 0.5 magnitudes deeper.

\begin{figure}
\centering
\includegraphics[width = .7\textwidth]{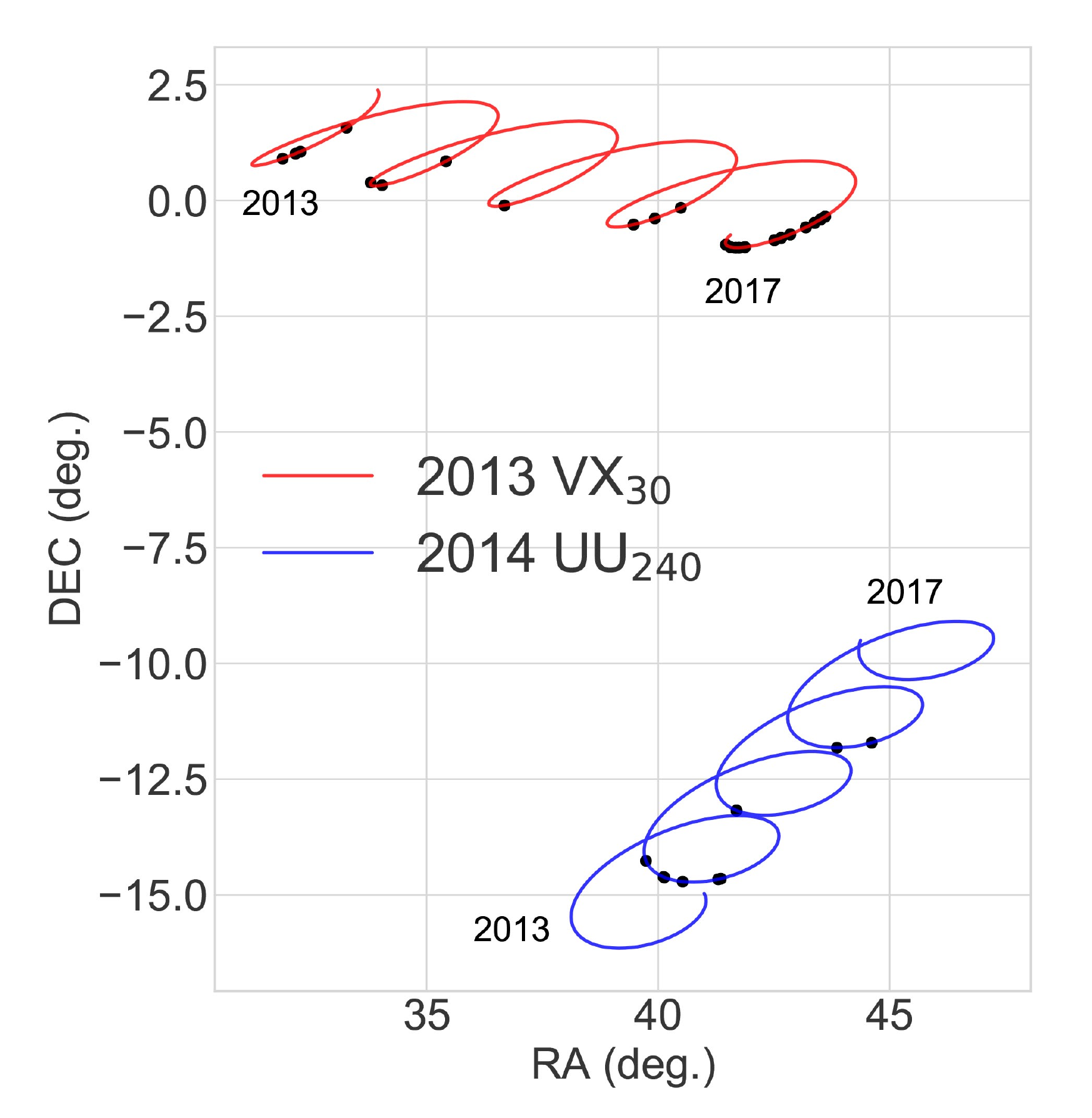}
\caption{Trajectories on the sky of \NTnamea\ and \NTnameb\ during the period 2013-2017. Dots indicate the epochs at which the objects
were observed by the DES. During the 2017-18 DES observing campaign, \NTnamea\ entered one of the supernova survey fields, providing a number of near-simultaneous observations in the $griz$ bands at approximately weekly intervals. \NTnamea\ and \NTnameb\ have observational arcs of 1566.8 days and 754.9 days respectively. \label{fig:discovery_arcs}}
\end{figure}

The new Trojans described in this work were discovered in DES wide survey data using search techniques similar to those described in \citet{Gerdes2017} and \citet{Becker2018}. A catalog of transient detections is created by applying the DES image-differencing pipeline \citep{Kessler2015} and artifact-rejection procedure \citep{Goldstein2015} to wide-survey images collected during the first four years of the survey. TNOs are identified by linking pairs of detections separated by 60 nights or less, whose separation is consistent in direction and apparent rate of motion with that expected from distant object subject to earth parallax. Once a 
preliminary discovery arc at one opposition has been obtained, it can readily be extended to include any additional detections at other oppositions.
Figure~\ref{fig:discovery_arcs} shows the trajectories on the sky of \NTnamea\ and \NTnameb\ during the period 2013-2017. \NTnamea\ was originally detected in DES wide survey exposures between Oct.\ 2013 and Dec.\ 2016. During the 2017-18 DES observing campaign, this object entered the shallow Stripe-82 supernova search fields, where it was observed in 44 exposures at 12 different epochs between Oct. 2017 and Jan. 2018. \NTnameb\ was observed in 13 wide survey exposures between Oct.\ 2015 and Dec.\ 2016.

To obtain the best possible orbit fit, we refined the original astrometric measurements using the \textsc{WCSfit} software described in \citet{Bernstein2017}. This code provides astrometric solutions using Gaia DR1 \citep{Gaia_DR1}. It also incorporates corrections for
tree-ring and edge distortions on the DECam CCDs, as well as for chromatic terms from lateral color and differential atmospheric refraction. Table~\ref{tab:elements} lists the barycentric osculating orbital elements of \NTnamea\ and \NTnameb\ at their respective discovery 
epochs. Both of them have orbital inclinations higher than 30 degrees, making them the two highest-inclination known stable Neptunian Trojans.  \NTnamea\ has absolute magnitude $H_r \sim 8.2$. Assuming \NTnamea\ is round and has an albedo of 0.05, it is roughly 140~km in diameter. \NTnameb\ may be slightly larger; it has $H_r \sim 8.0$ and could have a diameter of about 150~km.

\begin{table}
\caption{Barycentric osculating orbital elements and resonant dynamics \label{tab:elements}}
\begin{tabular}{lcc}
\hline
  & \NTnamea & \NTnameb \\
\hline
\hline
$N_{obs}$   & 60                       & 13         \\
arc (days)  & 1566.8                   & 745.9      \\
Epoch JD    & 2456567.7                & 2456959.83 \\
a (au)      & $30.08760  \pm 0.00061$  & $30.05716  \pm 0.00112$ \\
e           & $ 0.083744 \pm 0.000015$ & $0.048448  \pm 0.000133$ \\
inc (deg)   & $31.258593 \pm 0.000042$ & $35.744341 \pm 0.000147$ \\
$\omega$ (deg)& $215.446   \pm 0.0117$   & $73.175    \pm 0.123$\\
$\Omega$ (deg)& $192.538406\pm 0.000042$ & $81.998207 \pm 0.000330$ \\
Perihelion date (JD) & $2458596.05 \pm 1.64$ & $2477450.46 \pm 20.91$\\ 
absolute magnitude (H$_r$)& 8.21 $\pm$ 0.03         & 8.0 $\pm$ 0.1        \\
diameter (km)$^a$         & $\sim$ 140              & $\sim$ 150           \\

$\phi_{mean}$ (deg)$^{b, d}$ & $59.09^{+0.02}_{-0.02}$ & $57.08^{+0.32}_{-0.10}$ \\
$A_{\phi_{1:1}}$ (deg)$^{c, d}$     & $5.06^{+0.21}_{-0.19}$  & $2.60^{+0.32}_{-0.42} $ \\
mean lifetime (Gyr)   & $>$ 4.5                 & $>$ 4.5 \\
\hline
\end{tabular}
\\
$^a$ assuming albedo = 0.05.\\
$^b$ mean resonant angle.\\
$^c$ libration amplitude.\\
$^d$ calculated from the 1 Myr integrations, see Section~\ref{sec:dyna}.
\end{table}

\section{Dynamical Stability}
\label{sec:dyna}

Long-term dynamical stability is the key factor that separates the stable Trojans from recently-captured, and hence temporary, Trojans. The stable Trojans most likely belong to the primordial population. To understand the dynamical behavior of the newly discovered Trojans, we have performed an ensemble of numerical simulations. To start, we obtained the covariance matrix for the six orbital elements of each object by fitting all of the available DES detections across five years with the \texttt{orbfit} code of \citet{Bernstein2000}. The code was slightly modified to generate heliocentric (rather than barycentric) orbital elements and the corresponding covariance matrix was used as input for the subsequent numerical integrations. We first produced 1000 clones of the object based on the covariance matrix in order to sample the errors in all six orbital elements. These 1000 clones were then numerically integrated for 1 Gyr using the \texttt{Mercury6} integration package \citep{Chambers1999}. In order to reduce the integration time, we exclude the terrestrial planets from the simulations but incorporate their masses into the central body (the Sun). 

The resonant angle $\phi_{1:1}$ for the 1:1 commensurability was calculated for each of the clones. Specifically, the angle $\phi_{1:1}$ is given by 
\begin{equation}
\label{eq:res}
\phi_{1:1} = \lambda_N - \lambda_T,    
\end{equation}
where $\lambda_N$ and  $\lambda_T$ are the mean longitudes ($\lambda = M + \Omega + \omega$) of Neptune and the Trojan, respectively. Along with $\phi_{1:1}$, the mean resonant angle ($\phi_{mean}$) and amplitude ($A_{\phi}$) were also calculated. The angle $\phi_{mean}$ is obtained by averaging the resonant angle over the first 1 Myr of the integration, and the amplitude $A_{\phi}$ is determined by the half-peak RMS amplitude of the angular variations. This half-peak RMS amplitude is $A_\phi=\sqrt{2} \sigma_{\phi}$, where $\sigma_{\phi}$ is the standard deviation of the resonant angle $\phi_{1:1}$ during the first 1 Myr time interval. With the $\phi_{1:1}$, $\phi_{mean}$ and $A_{\phi}$ all calculated, we can then determine the lifetime of each clone as a Neptunian Trojan. We consider the clones to no longer be Trojans when their libration width becomes sufficiently large, $\mid\phi_{1:1} - \phi_{mean}\mid  > 60^{\circ}$.

\begin{figure}
\includegraphics[width = .5\textwidth]{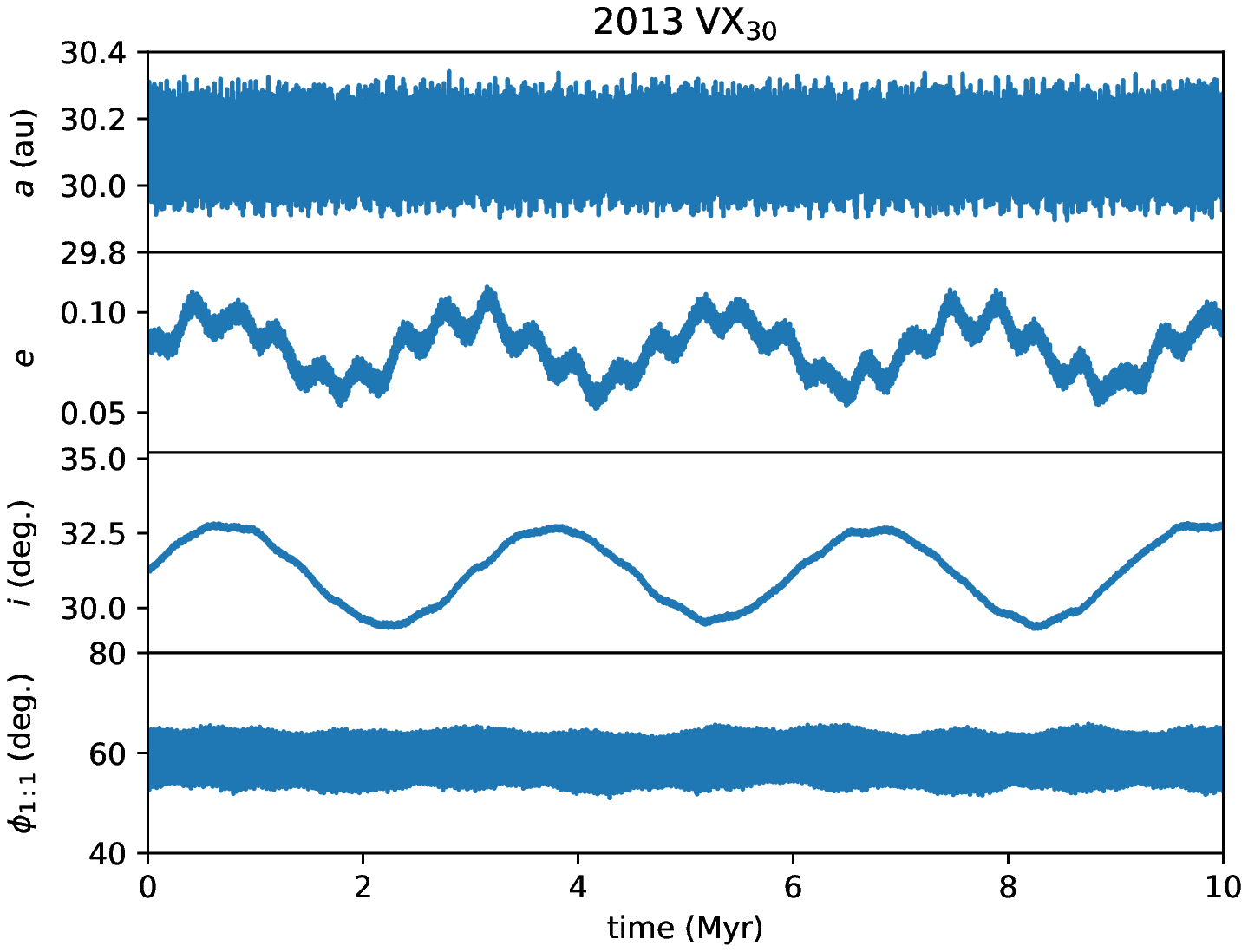}
\includegraphics[width = .5\textwidth]{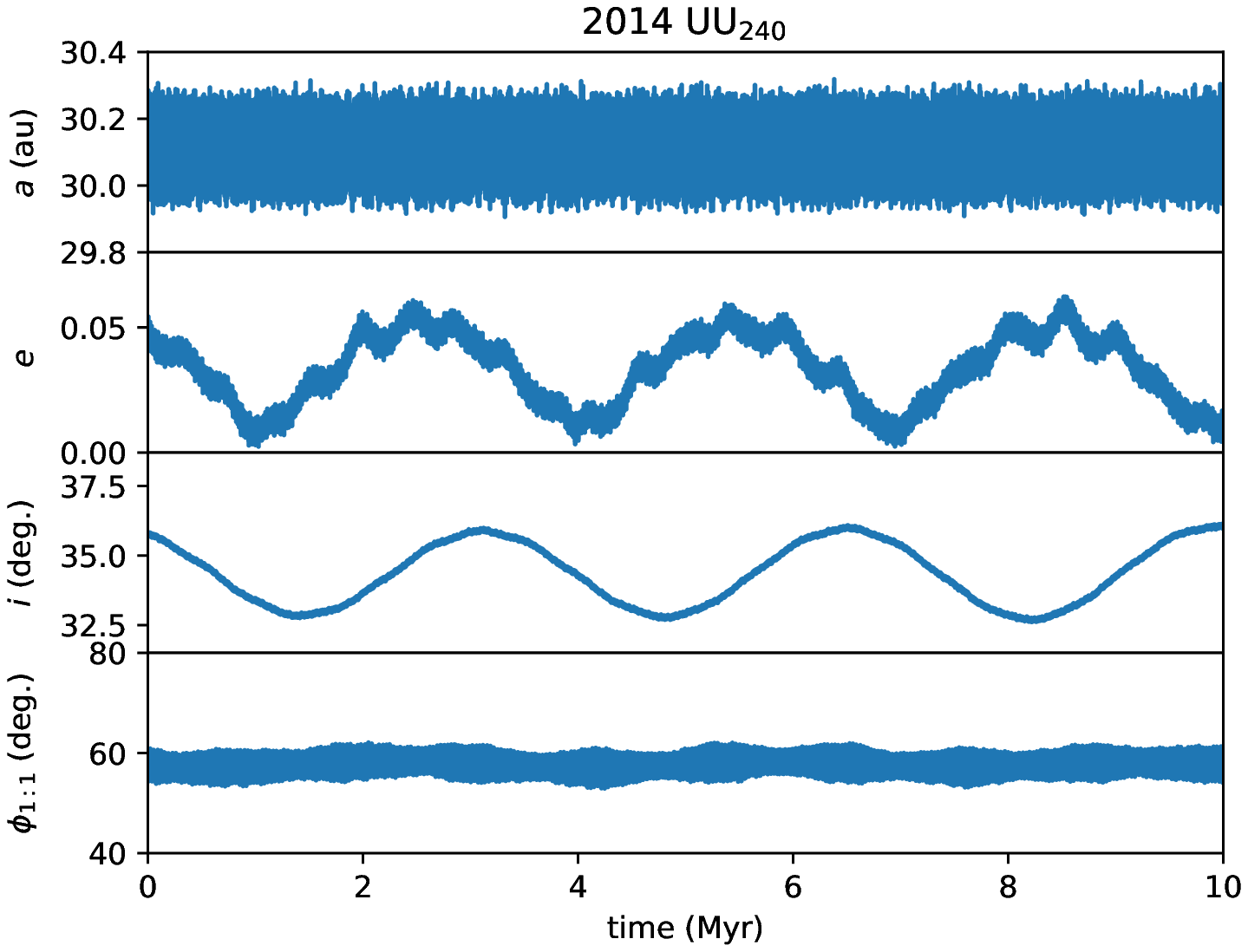}
\caption{Evolution of the orbital elements $a$, $e$, $i$, and the resonant angle $\phi_{1:1}$ over a 10 Myr time span for the best-fit orbits of \NTnamea\ (left) and \NTnameb\ (right). This pattern continues stably for nearly all of the 1000 clones analyzed, with all but three of \NTnamea's clones and all of \NTnameb's clones remaining in resonance over the entire 1 Gyr time span of the simulations. \label{fig: phi11}}
\end{figure}

These simulations show that both of the newly discovered Trojans are extremely stable. All of the clones of \NTnamea\ have a small libration amplitude ($A_{\phi} \sim 5^{\circ}$) and only three out of the initial 1000 clones  were lost during the 1~Gyr integration. The libration amplitude of \NTnameb\ is even smaller ($A_{\phi} \sim 2.6^{\circ}$), and none of the clones was lost before reaching 1~Gyr. Figure~\ref{fig: phi11} shows the three orbital elements of the two Trojans, along with resonant angle variations as a function of  time. These plots demonstrate the stability of these two Trojans. Because $99.7\%$ and $100\%$ clones of \NTnamea\ and \NTnameb, respectively, remain as Trojans for 1~Gyr, we conclude that both Trojans have lifetimes longer than the age of the Solar System and belong to the stable Trojan population. The dynamical properties of the newly discovered Trojans are listed in Table~\ref{tab:elements}.

\section{Photometry}
\label{sec:phot}

Three other L4 Neptunian Trojans discovered prior to this work have also been observed by the DES. Two of them, 2014~QO$_{441}$ and 2014~QP$_{441}$, were discovered by DES in the deep XMM-LSS supernova search field \citep{Gerdes2016} and were observed there during the 2014 and 2015 oppositions. The other Neptunian Trojan, 2011~SO$_{277}$, was discovered by the PS1 survey \citep{Lin2016} and was independently linked in 16 DES wide survey exposures between Sept. 2013 and Dec. 2017. 

Excluding the shallow-field supernova observations, the DES color observations of the five Neptunian Trojans are generally non-simultaneous. However, \NTnamea\ was observed in the shallow supernova field, and we were able to obtain simultaneous colors. For the objects that were observed in the deep supernova fields, we took the average $g$, $r$, $i$ and $z$ magnitude using the data taken in the deep fields only. For objects with wide-survey observations only, we average all of the available $g$, $r$, $i$ and $z$ measurements. We calculated the average absolute magnitude for each color band by normalizing the heliocentric and geocentric distance to 1~au and fitting a standard HG \citep{Bowell1989} phase curve with G=0.15\footnote{Evidence shows that the HG model may not well represent the phase curve of Jovian Trojans, because Jovian Trojans lack significant opposition surge \citep{Shevchenko2012}. However, many studies still apply the HG phase curve with G=0.15 as the standard to calculate the H magnitude of TNOs. We note that this issue may induce systemic error in H magnitude estimations but should not induce extra uncertainties in our color calculations.} to correct for brightness changes induced by the phase function. The observation phase angle of the Trojans ranged from 0.5 to 2 degrees. \textcolor{black}{The brightness variations of the Trojans are about 0.1 to 0.3 magnitude and possible due to the rotational effects.} We are unable to measure rotational periods for those Trojans, and therefore we do not attempt to account for the fact that our measurements may have been obtained at different points of rotational phase. However, the use of average magnitudes should remove, or at least average over, any possible rotational effects. 
Table~\ref{tab:color} shows the $r$-band magnitude, absolute magnitude, and the $(g-r)_{DES}$, $(r-i)_{DES}$ and $(i-z)_{DES}$ colors of the five Neptunian Trojans observed by DES. 
\NTnamea\ clearly has a much redder color than the other four. The $g-r$ color of 2011~SO$_{277}$ is not reliable, due to only three/two sparsely distributed $g$-/$r$-band observations. 

\begin{figure}
\centering
\includegraphics[width = .5\textwidth]{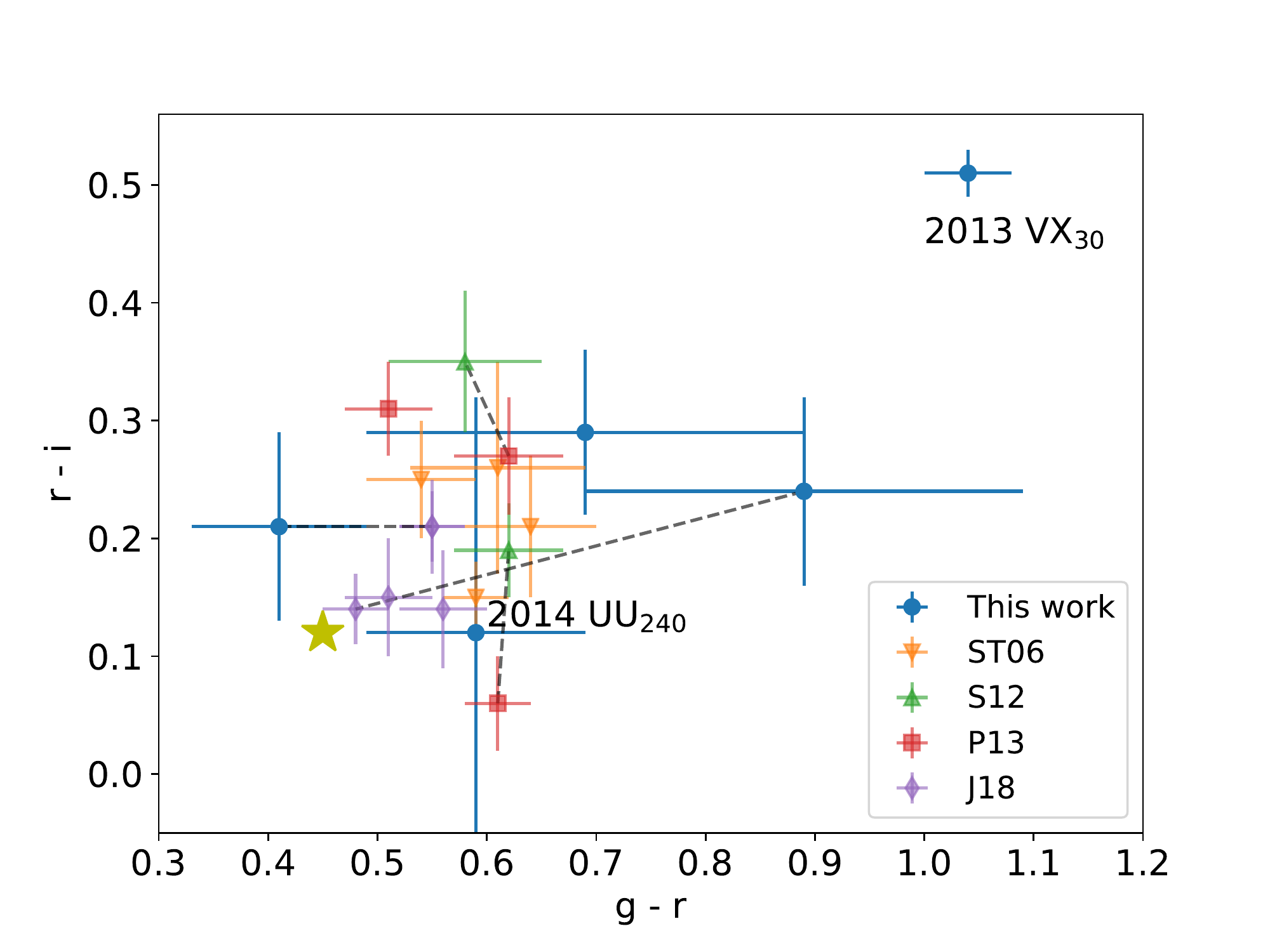}
\caption{Observed $r-i$ vs. $g-r$ colors of the Neptunian Trojans. \NTnamea\ stands out in the upper right portion of the diagram. The yellow star indicates the Solar color. The objects with multiple measurements are linked with dashed lines. All measurements are converted to the SDSS photometric system. The BVR measurements of J18 were converted to SDSS colors using the transformation equations of \citet{Fukugita1996} and \citet{Smith2002}. Reference: ST06\citep{Sheppard2006}, S12\citep{Sheppard2012}, P13\citep{Parker2013}, J18\citep{Jewitt2018}. \label{fig: color}}
\end{figure}

We compare colors of these DES-measured objects with other Neptunian Trojans with measured colors, as shown in Figure~\ref{fig: color}. The other photometry measurements were obtained by \citet{Sheppard2006} (ST06), \citet{Sheppard2012} (S12), \citet{Parker2013} (P13) and \citet{Jewitt2018} (J18). All measurements were converted into the SDSS photometry system \citep{DES_DR1, DES_Y1}. In such case: 
\begin{equation}
\label{eq:gr}
(g-r)_{SDSS} = (g-r)_{DES} - 0.01, \\
\end{equation}

\begin{equation}
\label{eq:ri}
(r-i)_{SDSS} = (r-i)_{DES} +0.069(g-r)_{DES} -0.25(i-z)_{DES}+ 0.02, \\
\end{equation}
and 
\begin{equation}
\label{eq:iz}
(i-z)_{SDSS} = 1.17(i-z)_{DES}+0.01.
\end{equation}
Four objects have been observed in two different studies, and all available measurements were plotted in Figure~\ref{fig: color}. The color of \NTnamea\ stands apart from all other known Neptunian Trojans and is ultra-red.  

\begin{figure}
\centering
\includegraphics[width = .5\textwidth]{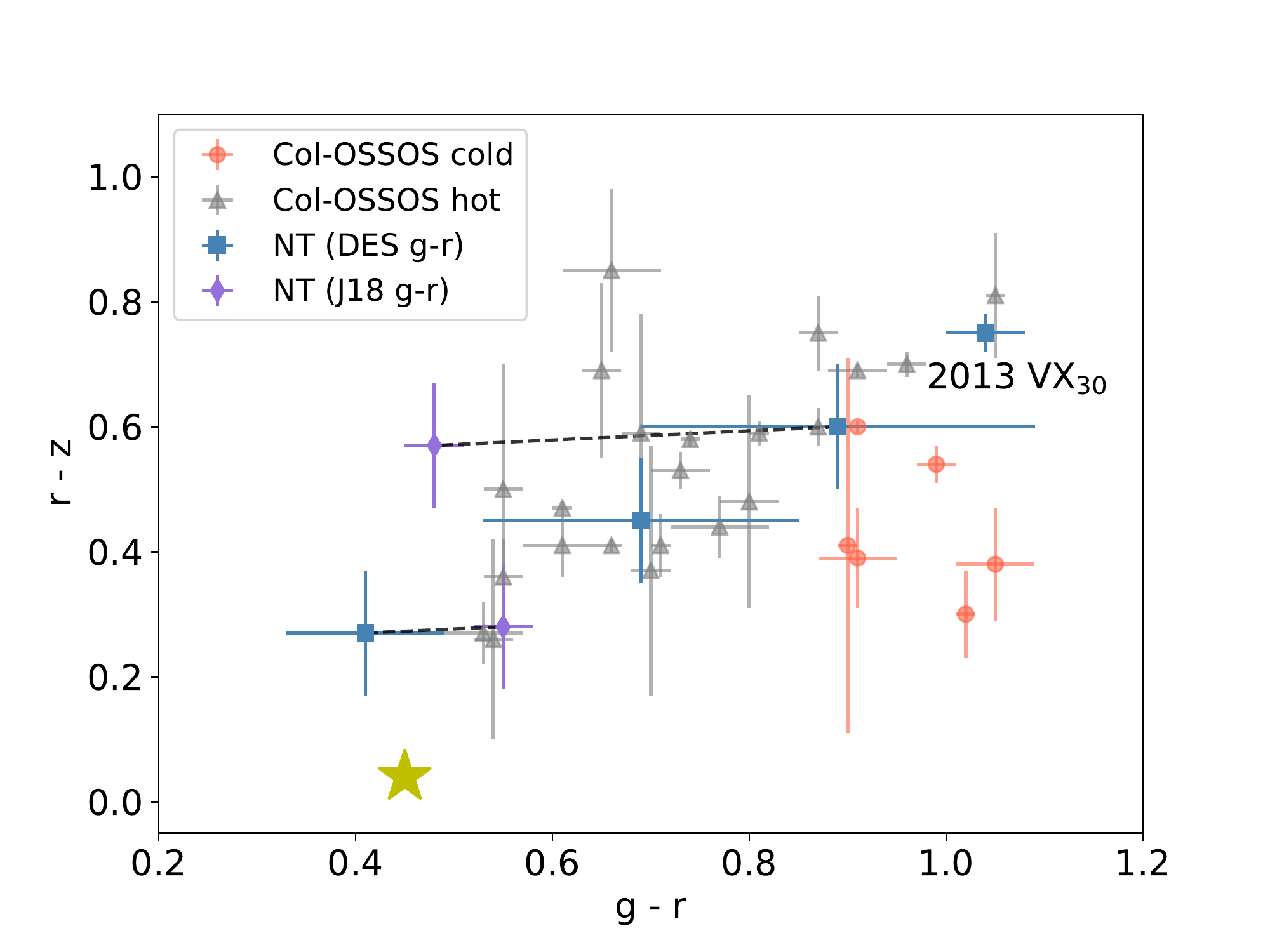}
\caption{Observed $r-z$ vs $g-r$ colors of the Neptunian Trojans (NT, blue squares and purple diamonds). Purple diamonds use $g-r$ of J18 (converted from $B-V$ color) instead of the DES measurements. The objects with multiple measurements are linked with dashed lines. The yellow star indicates the Solar color. The Col-OSSOS colors of the cold classical TNOs (Col-OSSOS cold, red circles) and dynamically excited TNOs (Col-OSSOS hot, gray triangles) were obtained from \citet{Pike2017}. \label{fig: grz}}
\end{figure}

The color of \NTnamea\ most closely resembles those of the ultra-red dynamically excited or cold classical TNOs. Figure~\ref{fig: grz} demonstrates the Neptunian Trojan $g-r$ and $r-z$ colors together with the Col-OSSOS TNO color sample \citep{Pike2017}. \NTnamea\ is ultra-red in $g-r$, and its $r-z$ color is also redder than the cold classical TNOs. It is closest in color to the dynamically-excited TNOs. \textcolor{black}{This result shows that except the cold classicals TNOs, all other TNO populations, include Neptune Trojans, are bi-modals in optical color.}

\begin{table}
\caption{Photometry \label{tab:color}}
\begin{tabular*}{\textwidth}{l @{\extracolsep{\fill}} cccccc}

\hline
Object & $N_{obs}$ & r & $H_r$ & $g-r$ & $r-i$ & $i-z$ \\
\hline
\hline
2011 SO$_{277}$ & 16 & 22.46 $\pm$ 0.03 & 7.55 &  0.9 $\pm$ 0.2$^a$  & 0.23  $\pm$ 0.08  & 0.30   $\pm$ 0.09 \\
2014 QO$_{441}$ & 164 & 23.46 $\pm$ 0.06 & 8.15 &  0.42 $\pm$ 0.08  & 0.17  $\pm$ 0.08  & 0.04  $\pm$ 0.07  \\ 
2014 QP$_{441}$ & 110 & 24.03  $\pm$ 0.06 & 9.26 &  0.7 $\pm$ 0.2  & 0.25 $\pm$ 0.07 & 0.13 $\pm$ 0.08 \\
\NTnamea        & 60 & 22.74 $\pm$ 0.03 & 8.21 & 1.05 $\pm$ 0.04 & 0.47 $\pm$ 0.02 & 0.20 $\pm$ 0.02 \\ 
\NTnameb        & 13 & 23.1  $\pm$ 0.1  & 8.0  &  0.6 $\pm$ 0.1  & 0.1  $\pm$ 0.2  & --              \\
\hline
\end{tabular*}
\\
$^a$ Not reliable. See section~\ref{sec:phot}.
\end{table}

\section{Blue and ultra-red Trojan Population Ratio}
\label{sec:pop}

In Sections~\ref{sec:dyna} and ~\ref{sec:phot}, we showed that \NTnamea\ is robustly dynamically stable, and ultra-red. The existence of a ultra-red Neptunian Trojan implies that the Neptunian Trojans contain both blue and ultra-red sub-populations. The apparent ratio of blue to ultra-red objects is 4:1 for the Neptunian Trojans observed by DES. In addition to problems associated with the extremely small sample size, however, the very different colors of blue and ultra-red Trojans may affect their detectability. The blue objects will be easier to detect in $g$-band, but difficult in $i$- and $z$-band, and vice-versa. To assess this effect, we use the DES Survey Simulator \citep{Hamilton2018} to examine whether the observed 4:1 ratio of blue to ultra-red Neptunian Trojans in DES data could be skewed by the effects of color-dependent detectability. 

To perform the survey simulation, we modeled the blue and ultra-red Trojan sub-populations with identical orbital and physical parameters, aside from their colors. We assigned the blue Trojans to have colors $g-r = 0.6$, $r-i = 0.2$ and $i-z = 0.1$. For the ultra-red Trojans, we used $g-r = 1.0$, $r-i = 0.5$ and $i-z = 0.2$. For the orbital parameters, the semi-major axes ($a$) were fixed at 30.1~au. The mean resonant angle  ($\phi_{mean}$) were fixed at the L4 Lagrange point, which is 60 degrees ahead of Neptune. The arguments of periapsis ($\omega$), longitudes of the ascending node ($\Omega$) and mean anomalies ($M$) were uniformly distributed and consistent with equation~\ref{eq:res}. The eccentricity ($e$) and libration amplitude ($A_{\phi_{1:1}}$) distributions were modeled by a Rayleigh distribution suggested by \citet{Parker2015}, which has the width as its only parameter. We used the suggested eccentricity width ($\sigma_e$) of 0.044 by \citet{Parker2015}, and the suggested libration amplitude width ($\sigma_{A_{\phi}}$) of 15 degrees by \citet{Lin2018}. The inclination ($i$) was modeled by a $\sin i$~$\times$~gaussian distribution \citep{Brown2001} with an inclination width ($\sigma_i$) of 26 degrees and truncated at 60 degrees, as suggested by \citet{Lin2018} for the hot component of the Neptunian Trojans. The absolute magnitude distribution was modeled by a power law distribution with a divot, which fits the dynamically excited TNO population well \citep{Lawler2018}. All of the modeling parameters are listed in Table~\ref{tab:model}.

\begin{table}
\caption{Neptunian Trojan Population Model \label{tab:model}}
\begin{tabular}{lcc}
\hline
Parameter & Distribution & Value \\
\hline
\hline
$a$ (au) & constant & 30.1 \\
$e$ & Rayleigh & width $\sigma_e$ = 0.044 \\
$i$  (degrees) & $\sin i \times$ gaussian \citep{Brown2001} & width $\sigma_i$ = 26$^a$ \\
$\omega$ (degrees) & uniform & 0 - 2$\pi$ \\
$\Omega$ (degrees) & uniform & 0 - 2$\pi$ \\
$M$ (degrees) & uniform & 0 - 2$\pi$ \\
$\phi_{mean}$ (degrees) & constant & 60 \\
$A_{\phi_{1:1}}$ (degrees) & Rayleigh & width $\sigma_{A_{\phi}}$ = 15 \\
$H$ distribution & power law with divot \citep{Lawler2018} & $\alpha_b$ = 0.9, $\alpha_f$=0.5, c=3.2, $H_b$=8.3$^b$\\  
$g-r$ & constant & 0.6 or 1.0 \\
$r-i$ & constant & 0.2 or 0.5 \\
$i-z$ & constant & 0.1 or 0.2 \\

\hline
\end{tabular}
\\
$^a$  truncated at $i = 60^{\circ}$.\\
$^b$ $\alpha_b$: bright end power law index, $\alpha_f$: faint end power law index, c: contrast between bright and faint end, H$_b$: break point. See \citet{Lawler2018} for details.
\end{table}

\begin{figure}
\centering
\includegraphics[width = .7\textwidth]{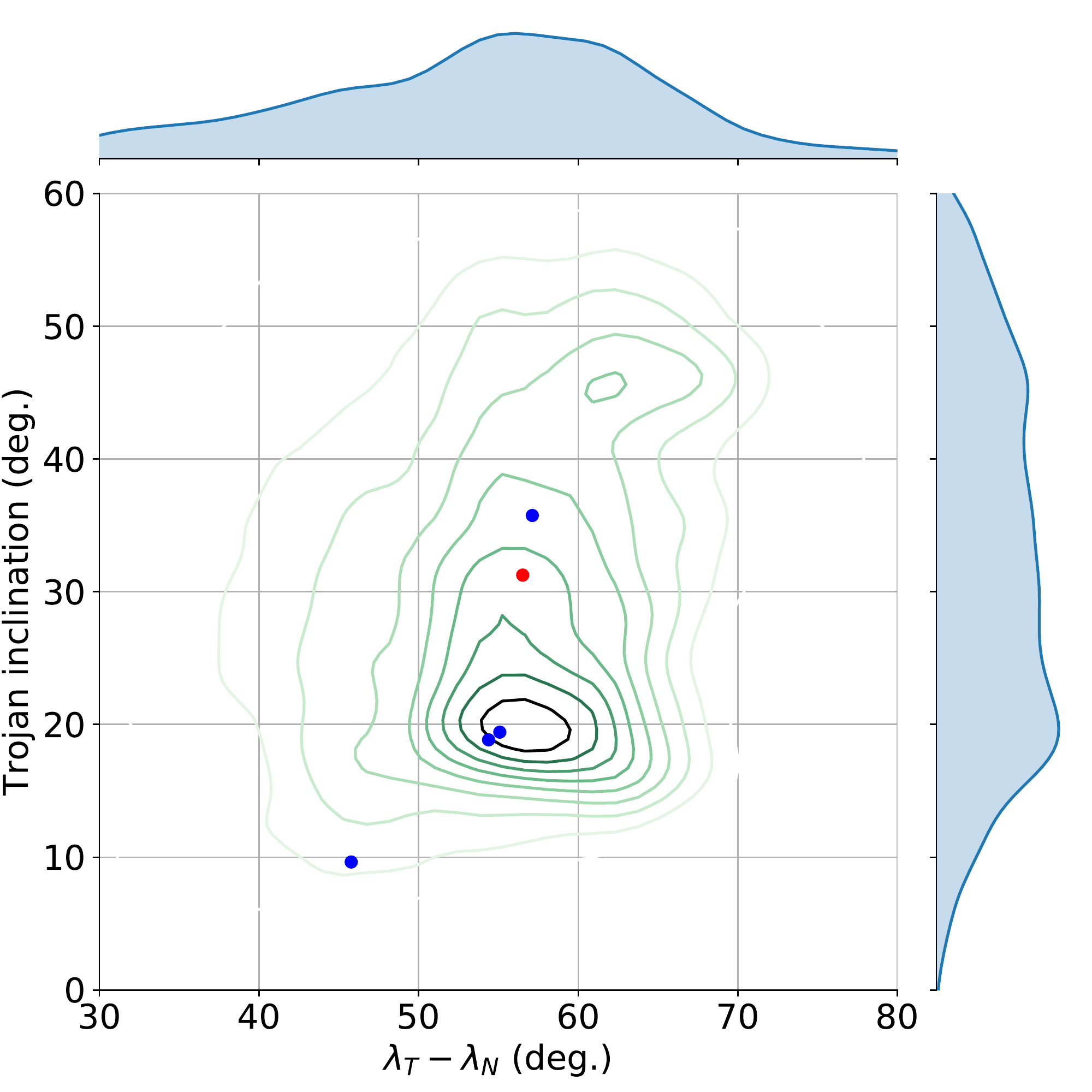}
\caption{Inclination vs. longitudinal separation from Neptune for synthetic
Neptunian Trojans detected with the DES survey simulator. The five Trojans observed in the data are indicated by dots, with \NTnamea\ in red and the others in blue. \label{fig:inc_vs_Nlon}}
\end{figure}

We injected 50,022 synthetic L4 Neptunian Trojans (24,934 blue, 25,088 red) with $H_r < 10$ into the Survey Simulator. The Survey Simulator produced a catalog of object detections that fell into a DES exposure and were brighter than that exposure's limiting magnitude. We then processed these detections through the DES moving object pipeline \citep{Gerdes2018} to estimate the actual detection rate. Of the total injected objects, 15,138 (7,513 blue, 7,625 red) objects fell into DES survey footprint, and 3,567 (1,579 blue, 1,988 red) of them were bright enough and had five or more detections to be detected by the pipeline. The pipeline detected total 1,541 synthetic L4 Neptunian Trojans (661 blue, 880 red), which is about $43\%$ of the detectable objects. Figure~\ref{fig:inc_vs_Nlon} shows inclination as a function of 
the angular separation from Neptune (in degrees of ecliptic longitude)
for simulated Trojans detected by our pipeline. The five Trojans detected in the data show good agreement with this distribution. 

Considering that the five Neptunian Trojans detected in the DES have $H_r < 10$, our result implies a total population (blue and red) of $162 \pm 73$ Neptunian Trojans with $H_r < 10$ at the L4 cloud. 
\citet{Sheppard2010:size_NT} estimated about 500 Neptunian Trojans with size $\gtrsim 50$ km. Assuming a 5$\%$ albedo, this corresponds to $H_r \sim 10$. If the L4 and L5 Neptunian Trojan have equal sized populations, that estimate would imply about 250 Neptunian Trojans at the L4 cloud. This result is consistent with our $162 \pm 73$ estimation. \textcolor{black}{Considering that the MPC (Minor Planet Center) lists 75 Jovian Trojans for $H_r <10$, our estimation of 320 Neptunian Trojans for both L4 and L5 camps with $H_r <10$ results about 4 times more population of Neptunian Trojans than Jovian Trojans.}

Moreover, \citet{Gladman2012} provided a upper limit of Neptunian Trojan population of $< 300$ at 95\% confidence for $H_g < 9.16$. Assuming $g-r = 0.6$, the upper limit will be $< 300$ for $H_r < 8.56$. There are total 12,891 objects with $H_r < 8.56$ in our synthetic L4 Neptunian Trojans, and our Survey Simulation detected 1,299 of them. Considering four of the five DES Neptunian Trojans have $H_r < 8.56$, we estimate that there are $40 \pm 20$ L4 Neptunian Trojans with $H_r < 8.56$. Again, assuming L4 and L5 Neptunian Trojan have equal populations, our estimation of 80 Neptunian Trojans for the L4 and L5 clouds together is below the upper limit of 300 suggested by \citet{Gladman2012} and  \textcolor{black}{should be in the order about $1\%$ of the Plutino (objects in a 3:2 mean motion resonance with Neptune) population ($\sim8000$ for $H_r < 8.66$ \citep{Volk2016})}.




%
%
%

Furthermore, due to the DES strategy of observing primarily in the redder bands ($riz$), the efficiency for detecting ultra-red Trojans is slightly higher, about 1.33 (880/661) times higher than for blue Trojans. (It
is not surprising that a survey whose primary scientific purpose is the study of the high-redshift universe should also excel at detecting red objects in our Solar System.)
Therefore, the corrected apparent blue-to-ultra-red Neptunian Trojan ratio is about 5:1. 
\textcolor{black}{If we further consider that ultra-red TNOs could have a higher average albedo than blue TNOs, which is $12\%$ and $6\%$ for ultra-red and blue, respectively \citep{Lacerda2014}, the detectability of ultra-red objects would be higher still. Similar argument and estimation have been done for the dynamically excited TNOs by \citet{Schwamb2018} and resulted a factor of 3.4 more blue objects then ultra-red ones comparing with the apparent blue-to-ultra-red object ratio. Thus, the true blue-to-ultra-red Neptunian Trojan population ratio should be 17:1.}

\section{Discussion}
\label{sec:dis}

Although the discovery of a ultra-red Neptunian Trojan makes the population more consistent with the idea that the TNOs act as the primordial reservoir, a problem remains. \textcolor{black}{With the roughly 17:1 ratio of blue to ultra-red objects, as described above, the number of ultra-red Neptunian Trojans is still too low, since the putative parent population (the TNOs\footnote{More specifically, the dynamically excited TNOs. The cold classical TNOs are mostly ultra-red.}) has blue and ultra-red objects ratio about 4.4 to 11.0 \citep{Schwamb2018}.}

\textcolor{black}{Collisional resurfacing is one potential mechanism to change the blue-to-ultra-red ratio. Here we consider a possible scenario: \citet{Almeida2009} argued that possible collisions between Neptunian Trojans and Plutinos could explain the observed size-inclination and color-inclination dependence of the Plutino population. They found that if Neptunian Trojans have population as many as Plutinos, the Trojan-Plutino encounters could be more frequent than the corresponding Plutino-Plutino encounters, and that the low-inclination Plutinos are more likely to collide with Neptunian Trojans than are high-inclination Plutinos. These results can explain why the sizes and colors of the Plutinos are a sensitive function of their orbital inclination. If this explanation holds for the Plutinos, then the size-inclination and color-inclination dependence should be also observed in the population of Neptunian Trojans. Indeed, recent evidence for size-inclination dependence in the Neptunian Trojan population \citep{Lin2018} may provide support for this Trojan-Plutino collision hypothesis. Moreover, we should also expect that the low-inclination Trojans have higher collision rates with Plutinos, and thereby tend to lose their primordial ultra-red colors. In this case, the original ultra-red surface should remain intact on some of the highly-inclined Neptunian Trojans. The newly discovered ultra-red object \NTnamea\ has the second-highest inclination of any known Neptunian Trojan. As a result, it is one of the Neptunian Trojans most likely to avoid collisions with Plutinos and thereby retain its primordial ultra-red surface.}

\textcolor{black}{However, there is a problem in this scenario: the population of Neptunian Trojans seem much less than the Plutino, at least in present-day. In section~\ref{sec:pop}, we estimated that the population of Neptunian Trojans is only $1\%$ of the Plutino population. Therefore, the Trojan-Plutino encounters should be less than Plutino-Plutino encounters and would not cause the size-inclination and color-inclination dependence in Plutino population. Regardless of Plutinos, will the number of Trojan-Plutino collisions be enough to change the blue-to-ultra-red ratio? We can roughly estimate the Neptunian Trojan collision rate as follow: \citet{Almeida2009} estimate the ratio of 1:1:5 for Neptunian Trojans-Neptunian Trojans (NT-NT), Neptunian Trojans-Plutino (NT-P) and Plutino-Plutino (P-P) collision rate based on Neptunian Trojans/Plutino ratio equal to 1/16. As we estimate that the Neptunian Trojans have $1\%$ of the Plutino population, the NT-NT, NT-P and P-P collision rate will be 1:6:30. The total population of Neptunian Trojans is about four times more than the Jovian Trojans. Assuming that the Neptunian Trojan clouds are 36 times more area than Jovian Trojan clouds ((30au/5au)$^2$), the number density of Neptunian Trojan will be 9 times lower than that of Jovian Trojan. Thus, the NT-NT collision rate would be $\sim 100$ times lower than Jovian Trojan collision. The NT-P collision rate is 6 times higher, therefore the collision rate in Neptunian Trojan clouds should be about $6\%$ of that in Jovian Trojan clouds. Considering the Jovian Trojans have at most a few major collisions, we consider that the number Neptunian Trojan collisions may not enough to change the blue-to-ultra-red ratio, if we give the present-day number of NT. It is possible that the Neptunian Trojans were more numerous before and had much higher collision rate to change the blue-to-ultra-red ratio, but such work needs to be done properly and beyond the scope of this paper.}

Nevertheless, the collision scenario outlined above has observable consequences: we expect the ratio of ultra-red to blue Trojans to be an increasing function of orbital inclination. Specifically, we should see significant numbers of ultra-red Trojans with inclination angles $i>30^\circ$. With these high inclination orbits, these Trojans should suffer fewer collisions and retain their original ultra-red surfaces. 

\section{Summary}
\label{sec:sum}

This paper reports the DES discoveries of two high-inclination Neptunian Trojans, \NTnamea\ and \NTnameb.  Our supporting numerical integrations indicate that both objects are dynamically stable over the lifetime of the Solar System, and likely belong to the primordial Neptunian Trojan population. \NTnamea\ and \NTnameb\ have orbital inclinations $i>30^{\circ}$, making them the highest-inclination stable Neptunian Trojans observed to date.  Our multi-color measurements of five Neptunian Trojans observed by the DES show that \NTnamea\ has significantly different colors than the other four; it has an ultra-red color similar to that of the reddest TNOs. As a result, we suggest that \NTnamea\ may be a ``missing link" between the Trojan and TNO populations.

With only one ultra-red Neptunian Trojan found, our survey simulations indicate that the ultra-red Neptunian Trojans may have a population 17 times smaller then their blue counterparts. This finding implies that the color distribution of Neptunian Trojans is still unlike their possible source, the TNO population, which has blue and ultra-red member ratio 4.4 to 11.

Based on \NTnamea's very high inclination and the argument that the low-inclination Plutinos are more likely to collide with Neptunian Trojans \citep{Almeida2009}, we propose that Trojan-Plutino collisions might be able to explain the observed surplus of blue Neptunian Trojans compared to ultra-red, \textcolor{black}{if the Neptunian Trojans once had much larger population than the present-day.} In this scenario, most of the lower-inclination Trojans were resurfaced by Trojan-Plutino collisions and thus have a blue color. The high inclination Neptunian Trojans, such as \NTnamea, are more likely to avoid collisions and retain their primordial ultra-red surfaces. 

Finally, we posited that if the Trojan-Plutino collision scenario is correct, the ratio of ultra-red to blue Trojans should be a sensitive function of orbital inclination. The LSST has the capability to detect and measure the colors of Neptunian Trojans over a wide range of inclinations, so that the ``Trojan Color Conundrum'' \citep{Jewitt2018} could be resolved within the next decade.

\bigskip
\noindent
{\bf Acknowledgment} 
\bigskip 

This material is based upon work supported by the National Aeronautics and Space
Administration under Grant No. NNX17AF21G issued through the SSO Planetary Astronomy Program, and by NSF grant AST-1515015.
S.J.H. is supported by the NSF Graduate Research Fellowship Grant No. DGE 1256260. 
This research has made use of data and services provided by the International Astronomical Union's Minor Planet Center.

Funding for the DES Projects has been provided by the U.S. Department of Energy, the U.S. National Science Foundation, the Ministry of Science and Education of Spain, 
the Science and Technology Facilities Council of the United Kingdom, the Higher Education Funding Council for England, the National Center for Supercomputing 
Applications at the University of Illinois at Urbana-Champaign, the Kavli Institute of Cosmological Physics at the University of Chicago, 
the Center for Cosmology and Astro-Particle Physics at the Ohio State University,
the Mitchell Institute for Fundamental Physics and Astronomy at Texas A\&M University, Financiadora de Estudos e Projetos, 
Funda{\c c}{\~a}o Carlos Chagas Filho de Amparo {\`a} Pesquisa do Estado do Rio de Janeiro, Conselho Nacional de Desenvolvimento Cient{\'i}fico e Tecnol{\'o}gico and 
the Minist{\'e}rio da Ci{\^e}ncia, Tecnologia e Inova{\c c}{\~a}o, the Deutsche Forschungsgemeinschaft and the Collaborating Institutions in the Dark Energy Survey. 

The Collaborating Institutions are Argonne National Laboratory, the University of California at Santa Cruz, the University of Cambridge, Centro de Investigaciones Energ{\'e}ticas, 
Medioambientales y Tecnol{\'o}gicas-Madrid, the University of Chicago, University College London, the DES-Brazil Consortium, the University of Edinburgh, 
the Eidgen{\"o}ssische Technische Hochschule (ETH) Z{\"u}rich, 
Fermi National Accelerator Laboratory, the University of Illinois at Urbana-Champaign, the Institut de Ci{\`e}ncies de l'Espai (IEEC/CSIC), 
the Institut de F{\'i}sica d'Altes Energies, Lawrence Berkeley National Laboratory, the Ludwig-Maximilians Universit{\"a}t M{\"u}nchen and the associated Excellence Cluster Universe, 
the University of Michigan, the National Optical Astronomy Observatory, the University of Nottingham, The Ohio State University, the University of Pennsylvania, the University of Portsmouth, 
SLAC National Accelerator Laboratory, Stanford University, the University of Sussex, Texas A\&M University, and the OzDES Membership Consortium.

Based in part on observations at Cerro Tololo Inter-American Observatory, National Optical Astronomy Observatory, which is operated by the Association of 
Universities for Research in Astronomy (AURA) under a cooperative agreement with the National Science Foundation.

The DES data management system is supported by the National Science Foundation under Grant Numbers AST-1138766 and AST-1536171.
The DES participants from Spanish institutions are partially supported by MINECO under grants AYA2015-71825, ESP2015-66861, FPA2015-68048, SEV-2016-0588, SEV-2016-0597, and MDM-2015-0509, 
some of which include ERDF funds from the European Union. IFAE is partially funded by the CERCA program of the Generalitat de Catalunya.
Research leading to these results has received funding from the European Research
Council under the European Union's Seventh Framework Program (FP7/2007-2013) including ERC grant agreements 240672, 291329, and 306478.
We  acknowledge support from the Australian Research Council Centre of Excellence for All-sky Astrophysics (CAASTRO), through project number CE110001020, and the Brazilian Instituto Nacional de Ci\^encia
e Tecnologia (INCT) e-Universe (CNPq grant 465376/2014-2).

This manuscript has been authored by Fermi Research Alliance, LLC under Contract No. DE-AC02-07CH11359 with the U.S. Department of Energy, Office of Science, Office of High Energy Physics. The United States Government retains and the publisher, by accepting the article for publication, acknowledges that the United States Government retains a non-exclusive, paid-up, irrevocable, world-wide license to publish or reproduce the published form of this manuscript, or allow others to do so, for United States Government purposes.



 \bibliographystyle{elsarticle-num-names}

\bibliography{references}

\end{document}